\documentclass[a4paper,11pt,reqno]{amsart}
\usepackage[T1]{fontenc}

\usepackage[numbers]{natbib}

\usepackage{dcolumn}
\usepackage{bm}

\usepackage[utf8]{inputenc}
\usepackage{mathptmx}

\usepackage{microtype}
\usepackage{amsmath,array}
\usepackage{textcomp}
\usepackage{hyperref}
\usepackage{mathptmx}
\usepackage{amssymb}
\usepackage{amsmath}
\usepackage{amsthm}
\usepackage{graphicx}
\usepackage{tabularx}
\usepackage{amsmath}
\usepackage{amsfonts}
\usepackage{amssymb}
\usepackage{graphicx}%
\usepackage{changes}

\providecommand{\U}[1]{\protect\rule{.1in}{.1in}}

\newcommand{\be}[1]{\begin{equation}\label{#1}}
\newcommand{\ee}{\end{equation}}

\let\pa\partial
\let\r\rho

\newtheorem{rem}{Remark}

\newcommand{\beq}{\begin{eqnarray}}
\newcommand{\eeq}{\end{eqnarray}}
\newcommand{\beqs}{\begin{eqnarray*}}
\newcommand{\eeqs}{\end{eqnarray*}}
\newcommand{\bequ}{\begin{equation}}
\newcommand{\eequ}{\end{equation}}

\def\r{\rho}

\def\r{\rho}

\newcommand{\eq}[1]{(\ref{#1})}

\newcommand{\ie}{\emph{i.e.}}
\newcommand{\rfr}[1]{#1_{\rm ref}}

\def\r{\rho}

\newcommand{\bfxi}{{\boldsymbol{\xi}}}

\newcommand{\cl}{c_{l}}
\newcommand{\cpd}{c_{pd}}
\newcommand{\Rd}{R_{d}}
\newcommand{\cpv}{c_{pv}}
\newcommand{\Rv}{R_{v}}

\newcommand{\joule}{\textnormal{J}}
\newcommand{\per}{\backslash}
\newcommand{\kilogram}{\textnormal{kg}}
\newcommand{\kelvin}{\textnormal{K}}

\newcommand{\DSI}{\textnormal{DSI}}
\newcommand{\DSIqg}{\textnormal{DSI}_{\textnormal{QG}}}
\newcommand{\DSIRo}{\textnormal{DSI}_{\textnormal{Ro}}}

\renewcommand{\S}{s}
\newcommand{\Sd}{\S_{(d)}}
\newcommand{\Bd}{B_{(d)}}
\newcommand{\Gd}{\mathbf{G}_{(d)}}
\newcommand{\Hd}{H_{(d)}}
\newcommand{\hd}{h_{(d)}}
\newcommand{\Pid}{\Pi_{(d)}}
\newcommand{\DSId}{\textnormal{DSI}_{(d)}}

\newcommand{\Sv}{\S_{(v)}}
\newcommand{\Bv}{B_{(v)}}
\newcommand{\Gv}{\mathbf{G}_{(v)}}
\newcommand{\Hv}{H_{(v)}}
\newcommand{\hv}{h_{(v)}}
\newcommand{\Piv}{\Pi_{(v)}}
\newcommand{\DSIv}{\textnormal{DSI}_{(v)}}

\newcommand{\Sc}{\S_{(c)}}
\newcommand{\Bc}{B_{(c)}}
\newcommand{\Gc}{\mathbf{G}_{(c)}}
\newcommand{\Hc}{H_{(c)}}
\newcommand{\hc}{h_{(c)}}
\newcommand{\Pic}{\Pi_{(c)}}
\newcommand{\DSIc}{\textnormal{DSI}_{(c)}}

\newcommand{\Sr}{\S_{(r)}}
\newcommand{\Br}{B_{(r)}}
\newcommand{\Gr}{\mathbf{G}_{(r)}}
\newcommand{\Hr}{H_{(r)}}

\newcommand{\Pir}{\Pi_{(r)}}
\newcommand{\DSIr}{\textnormal{DSI}_{(r)}}

\newcommand{\nsp}{n_{\rm sp}}
\newcommand{\Src}{Q}

\let\pa\partial

\newcommand{\changed}[1]{\textcolor{black}{#1}}

\begin{document}

\title{The dynamic state index with moisture and phase changes}



\author{S. Hittmeir}\address{Sabine Hittmeir: University of Vienna, Austria (sabine.hittmeir@univie.ac.at)} 
\author{R. Klein}\address{Rupert Klein: Freie Universit\"at Berlin, Germany (rupert.klein@math.fu-berlin.de)}
\author{A. M\"uller}\address{Annette M\"uller: Freie Universit\"at Berlin, Germany (annette.mueller@met.fu-berlin.de)}
\author{P. N{\'e}vir}\address{Peter N{\'e}vir: Freie Universit\"at Berlin, Germany (peter.nevir@met.fu-berlin.de)}

 %
 \date{\today}


\begin{abstract}

The dynamic state index (DSI) is a scalar field that combines variational information on the total energy and enstrophy of a flow field with the second law of thermodynamics. Its magnitude is a combined local measure for non-stationarity, diabaticity, and dissipation in the flow, and it has been shown to provide good qualitative indications for the onset and presence of precipitation and the organization of storms. The index has been derived thus far for ideal fluid models only, however, so that one may expect more detailed insights from a revised definition of the quantity that includes more complex aerothermodynamics.  The present paper suggests definitions of DSI-like indicators for flows of moist air with phase changes and precipitation. In this way, the DSI is generalized to signal deviations from a variety of different types of balanced states. A comparison of these indices evaluated with respect to one and the same flow field enables the user to test whether the flow internally balances any combination of the physical processes encoded in the generalized DSI-indices.

\end{abstract}

\maketitle 


\section{Introduction}


\subsection{The original DSI: Deviations from stationary, adiabatic, inviscid states in dry air}

The Dynamic State Index (DSI) is a parameter based on first principles of fluid mechanics that indicates local deviations of the atmospheric flow field from a stationary, adiabatic, and inviscid solution of the non-hydrostatic compressible governing equations \citep{Nevir2004}. In this way, the DSI can be evaluated on a given atmospheric flow field to detect atmospheric developments such as fronts or hurricanes. Atmospheric processes involve the interaction of energetic, thermodynamic, and vortex-related quantities. The (scalar) Dynamic State Index combines this information in a particular way. 

For dry conservative systems, the DSI is given by the Jacobi-determinant of three constitutive quantities: an advected scalar $\psi$, an energy variable $B$, here given by the Bernoulli stream function, and the potential vorticity (PV) $\Pi$:
\begin{equation} \label{eq:DSIgen}
\DSI=  \frac{\partial (\psi , B, \Pi)}{ \partial (a,b,c)} = \frac{1}{\rho} \frac{\partial (\psi , B, \Pi)}{ \partial (x,y,z)} \, ,
\end{equation}
where $x,y$ and $z$ denote the Cartesian coordinates \changed{ and $dm = da\ db \ dc$, where $a,b,c$ are the Lagrangian mass coordinates. Thus, integrating the conservation of mass in the Lagrangian sense with
\begin{equation} 
\rho =  \frac{\partial (a,b,c)}{ \partial (x,y,z)} \, ,
\end{equation}
leads to the right hand side representation of the DSI in \eq{eq:DSIgen} (See eq.~(8.2) in \citep{Nevir2004} with $\psi = s$ denoting the specific entropy)}. The precise definitions of $\psi$, $B $ and $\Pi $ depend on the underlying equations of motion. For more complex conservative systems involving moisture DSI-like indicators become sums of Jacobi-determinants as shown below. In the presence of irreversible processes, such as precipitation, such a compact representation turns out not to be available, however.

The formulation of the DSI in \eqref{eq:DSIgen} is equivalent to its representation as the mass flux divergence of the ``steady wind'' $\mathbf{v}_{st} = - (\nabla B \times \nabla \Psi) / \rho \Pi$  (c.f.\  \citet{Schaer1992}):
\beq
\DSI= -\frac{\Pi^2}{\rho} \nabla \cdot (\rho \mathbf{v}_{st}) = 0 \, .
\eeq
The field $\mathbf{v}_{st}$ may be interpreted as the local basic state wind that would have to prevail for given fields $(\psi, B, \Pi)$ for the flow to be stationary. Again, the precise definition of the basic state wind depends on the model equations. In the following $\mathbf{v}_{st}$ will simply be called steady wind. This principal definition of the DSI via the steady wind will be used below as a basis for the derivations of DSI-like indices for more complex systems involving moisture and precipitation. 

The basic state is characterized by DSI$=$0, and for dry dynamics this amounts to vanishing advection tendencies of the three constitutive quantities in the determinant \eqref{eq:DSIgen}, see, e.g., \citet{Sommer2009} or \citet{Mueller2019}. This property can be traced back to the Lagrangian conservation of the three constitutive quantities under the assumption of stationarity. The originally introduced DSI by \citet{Nevir2004} is based on the adiabatic non-hydrostatic compressible governing equations for dry air without consideration of thermodynamical sources and sinks, such as solar forcing, in the basic state. In this case, regarding \eqref{eq:DSIgen}, $\psi$ corresponds to the potential temperature, $B$ is the Bernoulli function (or total enthalpy) \citep[see][section 1.10]{Vallis2013} and $\Pi$ is Ertel's potential vorticity formed with the potential temperature as the advected scalar \citep[see][section 4.5]{Vallis2013}. 

This DSI concept can be applied to indicate non-steady, diabatic and frictional atmospheric processes across all scales: \citet{Weber2008} show how the characteristic dipole structure of the Dynamic State Index can be used to diagnose the evolution of high- and low-pressure areas on the synoptic scale, or hurricanes on the meso-scale. On the convective scale, several authors have shown that the DSI is strongly correlated with intensive precipitation processes, see e.g.\ \citet{Claussnitzer2008,Gassmann2014,Weijenborg2015}. 


\subsection{DSI-like indicators for generalized balances and fluids}

As stated above, the original Dynamic State Index generates a non-zero signal when the underlying flow is non-stationary or when it involves diabatic or dissipative processes. In other words, the DSI indicates deviations from a particular kind of balanced state. This perspective gives rise to generalizations of the DSI as indicators for deviations from different types of balance.

One prominent example in question is the geostropic and hydrostatic balance that forms the basis of quasi-geostrophic theory and is considered approximately valid on synoptic length and time scales. To accomodate such scale-dependent aspects in the \DSI\ framework, \citet{Mueller2018} derived a Dynamic State Index, $\DSIqg$, directly from the quasi-geostrophic model. In this case, the Bernoulli function reduces to $B = \frac{1}{f_0} \phi$, where $\phi$ is Earth's gravitational potential, \changed{$f_0$ denotes the Coriolis parameter} and $\Pi$ is the quasi-geostrophic, rather than the Ertel's, potential vorticity. By tracing back the original asymptotic derivation of the quasi-geostrophic model from the full compressible flow equations in \citep{Pedlosky1992}, these authors also showed that the $\DSIqg$ is the leading-order asymptotic approximation of the original $\DSI$ for compressible flows in this limit. As a consequence, by comparing the $\DSIqg$ with the original $\DSI$ for one and the same flow field of a compressible fluid, one can ascertain whether a balanced flow in the sense of vanishing $\DSI$ is also balanced in the quasi-geostrophic regime -- in which case the $\DSIqg$ should be comparably small. 

In the same spirit \citet{Mueller2019} recently introduced a $\DSI$ for the Rossby model and confirmed its high correlation to precipitation patterns and its applicability to the phenomenological concept of  ``Gro\ss wetterlagen''. For the (two-dimensional) Rossby model on the $\beta$-plane, the related $\DSIRo$ is given by the Jacobi-determinant of just two quantities, namely of the geopotential height ($B = \frac{1}{f_0} \phi$) and the absolute vorticity ($\Pi   = \zeta_a$). Utilizing the three indices $\DSI$, $\DSIqg$, and $\DSIRo$ available, one can now test whether a balanced flow ($\DSI \ll 1$) is also balanced on synoptic ($\DSIqg \ll 1$), or even on the scale of the external Rossby radius ($\DSIRo \ll 1$) and beyond.

Another possible $\DSI$ generalization consists of replacing the reference thermodynamic state change underlying the construction of the index from the isentropic ($p \sim \rho^\gamma$) to some polytropic ($p \sim \rho^\kappa$) pressure-density relationship, where $\gamma$ and $\kappa$ are the isentropic and polytropic exponents, respectively. With a change of this type, a non-vanishing $\DSI^\kappa$ can indicate, e.g., deviations from an isothermal, isochoric, or isobaric state for $\kappa = \gamma$, $1$, or $\infty$, respectively. Furthermore, minimization of \DSI$^\kappa$ with respect to $\kappa$ for a given slowly varying flow field characterizes the diabatic effects in the flow in terms of the most similar polytropic process. 

The derivation of a polytropic $\DSI$-family will be the subject of a forthcoming paper, but analogous considerations motivate the present work: The comparison, for one and the same flow field, of $\DSI$-like variables that encode balances under different prevailing moist processes provides diagnostic insights into the aerothermodynamic nature of the flow. Mathematically, in developing a $\DSI$ for moist processes, we consider a basis of constituting quantities that generalize the potential temperature, $\psi$, the Bernoulli function, $B$, and the potential vorticity, $\Pi$, for model equations that include the effects of moist processes.

After a brief summary of the mathematical formalism underlying the original $\DSI$ concept, a generalized derivation of the same quantity that directly invokes the second law of thermodynamics will be presented in section \ref{sec:DSIGeneralization}. Section \ref{sec:MoistDSIVariants} then utilizes the generalized derivation to provide a hierarchy of three $\DSI$-like indices relevant for moist air flows. These indices signal balances for moist flows with and without phase changes and for precipitating and non-precipitating states. Section~\ref{sec:Concl} summarizes our results and provides further conclusions. 


\section{Generalized derivation of DSI-like variables}
\label{sec:DSIGeneralization}


\subsection{Classical derivation of the DSI}
\label{subsec:DSIpe}

Here we recall the derivation of the DSI for the equations for non-hydrostatic compressible   
flows of dry air \citep[see][]{Nevir2004,Seltz2010} for later reference. We start with the momentum equation
\beq\label{eq.v.intro}
\pa_t\mathbf{v} + \boldsymbol{\xi}_a \times \mathbf{v}+ \Big(\frac12\boldsymbol{\nabla}\mathbf{v}^2  + \frac{1}{\rho} \boldsymbol{\nabla}p + \boldsymbol{\nabla} \phi \Big) =\mathbf{F}\,,
\eeq
where $\mathbf{F}$ denotes frictional forces, $\phi$ the geopotential height field, $p$ the pressure, $\rho$ the density and $\boldsymbol{\Omega}$ the earth rotation, and where we have utilized the Weber-Transform
\beq \label{eq:WeberTrafo}
\mathbf{v}\cdot \nabla \mathbf{v}= \bfxi \times \mathbf{v} + \frac12 \nabla  \mathbf{v}^2 \,,
\eeq
where $\bfxi= \nabla \times \mathbf{v}$ and $\bfxi_a = \bfxi + 2\boldsymbol{\Omega}$ denote the relative and absolute vorticities, respectively. Under adiabatic conditions the potential temperature
\beq\label{theta}
\theta= T\left(\frac{p_0}{p}\right)^{\frac{R_d}{c_{pd}}}\,
\eeq
is conserved along Lagrangian paths. Here $R_d$ and $c_{pd}$ are the ideal gas constant and the specific heat capacity at constant pressure for dry air, respectively.  Thus, forming the cross product of \eqref{eq.v.intro} with $\nabla \theta$, we obtain 
\begin{equation}\label{eq.v.ref}
\begin{split}
& (\pa_t \mathbf{v} - \mathbf{F} + \nabla B + \frac{1}{\rho} \nabla p - \nabla H) \times \nabla \theta \\
&= - (\bfxi_a\times \mathbf{v})\times \nabla \theta \ = \ \bfxi_a \, \mathbf{v}\cdot \nabla \theta - \mathbf{v} \, \bfxi_a\cdot\nabla\theta \\
&=- \bfxi_a \pa_t \theta- \mathbf{v} \, \r  \Pi^\theta , 
\end{split}
\end{equation}
where
\beq  \label{eq:PVpe}
B = \frac12 \mathbf{v}^2 + H +\phi\,,  
\quad H = c_{pd} T\,, 
\quad\textnormal{and}\quad \Pi^\theta = \frac{\bfxi_a\cdot\nabla \theta}{\r} \ \ 
\eeq
are the Bernoulli function, the enthalpy, and the potential vorticity, respectively. Using the ideal gas law
$p=R_d \r T$ we further obtain
\begin{equation}
\begin{split}
\left( \frac{1}{\rho} \nabla p - \nabla H\right) \times \nabla \theta & = T(R_d\nabla \ln p - c_{pd}\nabla \ln T)\times \nabla \theta \\ 
& = - c_{pd}T \nabla \ln \theta \times \nabla \theta =0 \,. 
\end{split}
\end{equation}
Assuming stationarity and neglecting friction leads to the steady wind for adiabatic, inviscid, and steady flows \citep{Schaer1992}
\beq
\mathbf{v}_{st}=- \frac{1}{\rho \Pi^\theta}\Big[ \nabla B  \times \nabla \theta \Big]\,.
\eeq
The DSI is designed to signal deviations from this steady wind, and its mathematical representation follows from the continuity equation based on the steady wind: 
\begin{equation}
\label{DSI0}
\begin{split}
\DSI 
& = -\frac{{\Pi^\theta}^2}{\rho} \nabla \cdot (\rho \mathbf{v}_{st}) 
= - \frac{1}{\rho}\nabla\Pi^\theta\cdot \Bigl(\nabla B \times \nabla \theta\Bigr) \\
& = \frac{1}{\rho} \frac{\partial(\theta,B,\Pi^\theta)}{\partial(x,y,z)}\,.
\end{split}
\end{equation}
Thus, the DSI indicates local deviations of an atmospheric flow field from a stationary, adiabatic, and inviscid state, i.e., the presence of instationary, viscous, or diabatic processes. 

Interpreting the DSI geometrically and regarding isentropic surfaces, the steady wind based on the  non-hydrostatic compressible governing  equations blows along the isolines of the Bernoulli stream function as well as of the PV, and the DSI signal indicates non-alignment of these fields in the sense that for non-zero $\DSI$ the vectors $\nabla \theta, \nabla B$, and $\nabla \Pi^\theta$ are linearly independent. In turn, such non-alignment signals the presence of molecular transport or more general diabatic atmospheric processes.

Previous works corroborate, on the basis of meteorological observations, that the DSI as defined in \eqref{DSI0} signals diabatic processes \citep{Claussnitzer2008,Gassmann2014,Weijenborg2015}. To obtain DSI variants that locate specific diabatic processes, for example the formation of clouds, or extreme precipitation, we will include successively more complex moist processes in the equations of motion and derive related new $\DSI$-like indices in section~\ref{sec:MoistDSIVariants}.
 

\subsection{DSI and PV for a multi-component fluid}
\label{subsec:GenDSI}


\subsubsection{A generalization of the $\DSI$}

To generalize the derivation of the DSI for a multi-component fluid we proceed in analogy to the last section. Starting with the corresponding set of equations of motion,  we derive the model-dependent steady wind leading to the scalar DSI-field that indicates deviations from this basic state. The point of departure for the derivations are again the momentum equations which we rewrite here as 
\beq \label{eq.v}
\pa_t\mathbf{v} + \boldsymbol{\xi}_a \times \mathbf{v}+ \nabla B = \mathbf{F} + \mathbf{G}\,,
\eeq
where 
\beq \label{eq.B}
B = \frac12 \mathbf{v}^2 + H +\phi
\eeq
and
\beq\label{eq.G}
\mathbf{G} = \nabla H - \frac{1}{\rho}\nabla p = T\nabla S + \sum_{i=1}^{\nsp} \mu_i \nabla Y_i
\eeq
are again the Bernoulli function (or total enthalpy) from \eq{eq:PVpe} with an appropriately scaled specific enthalpy $H$ adapted to the system under consideration, $G$ is the effective diabatic forcing term, $S$ the specific entropy, and $\mu_i = H_i - T\S_i$ and $Y_i$ are the chemical potential and the mass fraction of the $i$th fluid constituent \changed{with $i \in \{1,\dots, \nsp\}$ and $\nsp$ denoting the number of fluid components}. Let 
\beq \label{eq:PVgen}
\Pi^\S = \frac{\bfxi_a\cdot\nabla \S}{\r} \,,
\eeq
denote the Ertel-type potential vorticity based on the entropy $\S$ which satisfies the transport equation 
\beq\label{eq:EntropyEvolution}
\frac{d\S}{dt}= \pa_t\S +\mathbf{v}\cdot \nabla \S = \Src_\S
\eeq
with the generalized source term $\Src_\S$ that subsumes all dissipative and external forcing processes that affect entropy evolution along fluid path lines. 
Taking the cross product of \eqref{eq.v} with $\nabla \S$ yields 
\begin{equation} \label{eq.v.ref}
\begin{split}
(\pa_t \mathbf{v} + \nabla B - \mathbf{F} - \mathbf{G})\times \nabla \S
  & =  - (\bfxi_a\times \mathbf{v})\times \nabla \S \\
     & = \ \bfxi_a \, \mathbf{v}\cdot \nabla \S - \mathbf{v} \, \bfxi_a\cdot\nabla\S 
      \\ 
  & =  - \bfxi_a \pa_t \S - \mathbf{v} \, \r  \Pi^\S + \bfxi_a\, \Src_\S\,.
 \end{split}
\end{equation}
Following Schär's procedure for the case of dry air \citep{Schaer1992}, we now define a steady wind based on \eqref{eq.v.ref} by assuming stationarity ($\partial_t \equiv 0$). This yields 
\beq
\mathbf{v}^\S_{st}=- \frac{1}{\rho \Pi^\S}\Big[ (\nabla B - \mathbf{G} - \mathbf{F})\times \nabla \S - \bfxi_a\,  \Src_\S\Big]\,.
\eeq
The definition of the DSI according to \citet{Nevir2004,Sommer2009} then results from the continuity equation for the steady wind with a suitable normalization,
\begin{equation}
\begin{split} \label{eq.DSI}
\DSI^\S &= -\frac{(\Pi^\S)^2}{\rho} \nabla \cdot (\rho \mathbf{v}^\S_{st}) \\
& = \frac{(\Pi^\S)^2}{\rho}\nabla\cdot\Big[ \frac{1}{\Pi^\S} \Big( (\nabla B  - \mathbf{G} - \mathbf{F})\times \nabla \S - \bfxi_a\, \Src_\S
\Big)\Big]\,.
 \end{split}
\end{equation}
For a steady flow the $\DSI^\S$ vanishes \changed{by definition of the steady wind}, because under these conditions the steady wind coincides with the actual wind field and satisfies the continuity equation for steady conditions which amounts to 
\beq
\nabla \cdot (\r \mathbf{v}^\S_{st})=0 \quad \Leftrightarrow \quad \DSI^\S  = 0\,.
\eeq
Note that we have generalized the original $\DSI$ concept by allowing for both frictional forces, $\mathbf{F}$, and entropy production, $\Src_\S$. Of course, when $\mathbf{F} \equiv 0$ and $\Src_\S \equiv 0$ the indicator is tuned again to adiabatic, frictionless and steady flows as discussed in section~\ref{subsec:DSIpe}.

The generalized $\DSI$-concept as represented by \eq{eq.DSI} will be utilized below to suggest dynamic state indices for the equations of moist air flow. Entropy (or potential temperature), Bernoulli function, and PV will be adapted to the given model, leading to the formulation of the corresponding steady wind and of a related $\DSI$.


\subsubsection{Is there a generalization of the potential vorticity for inclusion in the DSI framework?}

The vorticity transport equation, and with it the Lagrangian evolution of the entropy-based potential vorticity $\Pi^{\S}$, can be derived as follows. Again, the point of departure are \eq{eq.v} and \changed{\eq{eq.G}} which we combine to yield
\beq \label{eq:dervor}
\frac{\partial \mathbf{v}}{\partial t} + \boldsymbol{\xi}_a \times \mathbf{v} + {\nabla} B
= T {\nabla}\S + \sum_{i=1}^{\nsp}  \mu_i \nabla Y_i + \mathbf{F} \, .
\eeq
Taking the curl yields the vorticity transport equation
\beq
\frac{\partial \boldsymbol{\xi}}{\partial t} + {\nabla}\times \left( \boldsymbol{\xi}_a \times \mathbf{v} \right) = {\nabla}T \times {\nabla}\S + \sum_{i=1}^{\nsp} \nabla \mu_i \times \nabla Y_i + \nabla\times\mathbf{F}\,. \ \ \ \ \ \ 
\eeq
Incorporating the conservation of mass via the continuity equation and multiplying by $\nabla \S$, omitting the term $(\nabla T\times\nabla\S) \cdot \nabla \S$, and utilizing \eq{eq:EntropyEvolution}, leads to the evolution equation for the entropy-based potential vorticity $\Pi^\S = (\boldsymbol{\xi}_a\cdot  \nabla \S) / \r$,   
\beq
\frac{d \Pi^\S}{dt} 
= \frac{1}{\rho} 
  \left(\sum_{i=1}^{\nsp} \nabla\mu_i\times\nabla Y_i + \nabla\times\mathbf{F}
  \right) \cdot \nabla \S + \frac{\boldsymbol{\xi}_a }{\rho} \cdot \nabla \Src_\S \,, \ \
\eeq 
where, following \citet{GassmannHerzog2015},
\beq
\nabla \mu_i = \frac{1}{\rho_i} \nabla p_i - \S_i \nabla T\,.
\eeq
Clearly, for frictionless, $\mathbf{F} \equiv 0$, and adiabatic, $\Src_\S \equiv 0$, flow we must require in addition that the solenoidal first terms on the right hand side vanish as well \citep[see e.g.][]{Schubert2004}.

With a more general advected scalar, $\psi$, used instead of entropy in forming the PV and $\DSI$ variables, $\Pi^\psi$ and $\DSI^\psi$, respectively, one would require 
\beq\label{cons.Pi}
\mathbf{F} = 0\,, \quad
\frac{d \psi}{dt}= \Src_\psi = 0\,,
\eeq
and
\beq
\nabla \psi \cdot\left(\nabla T \times \nabla \S + \sum_{i=1}^{\nsp} \nabla \mu_i \times \nabla Y_i \right) = 0
\eeq
for $\Pi^\psi$ to be a Lagrangian conserved quantity.

Analogous potential vorticity equations can also be derived for models including moist processes. But, in general, retaining $\rho$ as the total density and using the entropy~$S$ for the appropriate multi-species atmospheric models with moisture, we are not able to find a scalar $\psi$ such that the solenoidal term vanishes in all cases. \citet{Schubert2004} discusses the evolution of potential vorticities, $\Pi^{\theta_x}$, formulated in terms of different potential temperature-like variables $\theta_x$. He demonstrates that the solenoidal term always vanishes when using the virtual potential temperature, $\theta_\rho$, which is effectively a function of density and pressure only. This ansatz hides the influences of particular moist phase conversions, however, and this is why the Gibb's form involving $T\nabla S$ rather than $\nabla p/\rho$ is used in the momentum equation \eqref{eq:dervor} here, and why the subsequent derivations of DSI-like quantities are based on entropy rather than on potential temperatures.
 

\subsection{The DSI for dry air}
\label{subsec:DSIdry}

In the following the steady wind and the DSI for the non-hydrostatic compressible governing equations for dry air are derived corroborating the DSI introduced by \citet{Nevir2004}. For the derivations of the DSI for moist air we follow the same steps, but adapt the entropy, the Bernoulli function and the potential vorticity to the particular models.

\changed{Owing to the second law of thermodynamics, entropy is conserved along particle paths for adiabatic motions, and for dry air it satisfies the relation
\beq\label{ds.dry}
\frac{d \Sd}{dt} = c_{pd} \, \frac{d \ln T}{dt} - R_d\, \frac{d \ln p}{dt}\, . 
\eeq 
Integrated, this yields the entropy equation of state, 
\beq\label{s.dry}
\Sd= c_{pd}\, \ln \frac{T}{T_0} - R_d \, \ln \frac{p}{p_0}\,,
\eeq
where $c_{pd}$ denotes the specific heat capacity at constant pressure for dry air, see also table\,\ref{table}. We remark that the commonly used notation of $S$ for the entropy is changed here to $\Sd$ indicating that we consider the entropy for dry air. This will allow us to distinguish between the different entropies for dry air $(d)$, moist air with vapor only $(v)$, cloudy air $(c)$ and precipitating air $(r)$ later on. The other quantities will be indexed accordingly.}

\changed{With the entropy being a conserved quantity for adiabatic motion, a natural choice for $\psi$ is}
\beq
\psi= \Sd\,,
\eeq  
and, due to \eqref{cons.Pi}, the associated potential vorticity
\beq
\Pid^{\S} = \frac{\bfxi_a\cdot \nabla \Sd}{\rho}\,
\eeq
is also conserved during frictionless motion. 
 
\begin{table*}
	\caption{Thermodynamic equation of state parameters for moist air at reference temperature $T_0=273.15$ K and reference pressure $p_0 = 10^5 Pa$ (see, e.g., \cite{Cotton2010,Emanuel1994}):}
		\begin{tabular}{|crl@{\quad}l|}
			\hline \hline
			$\cpd$
			& $1005$
			& $\joule\per\kilogram\per\kelvin$
			& dry air specific heat capacity at constant pressure 
			\\
			$\Rd$
			& $287$
			& $\joule\per\kilogram\per\kelvin$
			& dry air gas constant
			\\
			$\cpv$
			& $1850$
			& $\joule\per\kilogram\per\kelvin$
			& water vapor specific heat capacity at constant pressure 
			\\
			$\Rv$
			& $462$
			& $\joule\per\kilogram\per\kelvin$
			& water vapor gas constant
			\\
			$\cl$
			& $4218$
			& $\joule\per\kilogram\per\kelvin$
			& liquid water specific heat capacity  
			\\
			$\rfr{L}$
			& $2.5\cdot 10^{6}$
			& $\joule\per\kilogram$
			& latent heat of condensation at reference conditions
			\\
			\hline \hline 
		\end{tabular}
	\label{table}
\end{table*}

For the dry air case the enthalpy $\hd$ (up to integration constants) is defined by 
\beq
\hd=c_{pd} T\,.
\eeq
Using the ideal gas law
\beq\label{ideal.d}
p=R_d \r T
\eeq
it follows that 
\beq\label{dh}
d \hd = T d\Sd +\frac{1}{\rho}dp\,.
\eeq
Thus setting $\Hd = \hd$ in the Bernoulli function in \eqref{eq.B} we obtain for $\Gd$ according to \eqref{eq.G} the identity
\beq
\label{G.dry} 
\Gd= \changed{ T\nabla \Sd}\,,
\eeq
which satisfies, obviously, $\Gd\times\nabla \Sd=0\,$ and
the steady wind is thus given by
\beq
\mathbf{v}^{\S}_{st (d)}=\frac{1}{\rho \Pid^{\S}}  \nabla \Sd \times \nabla \Bd\,.
\eeq
Then the DSI according to \changed{\eqref{eq.DSI}} reduces to 
\beq \label{eq:DSI_d}
\DSId^{\S}=\frac{1}{\rho}\nabla \Pid^{\S} \cdot(\nabla \Sd \times \nabla \Bd)\,,
\eeq
where we have used the fact that $\nabla \cdot (\nabla \Sd \times \nabla \Bd)=0$. Following \citet{Nevir2004} the DSI can also be expressed as
\beq
\DSId^{\S} =\frac{1}{\rho}\frac{\pa(\Sd,\changed{\Bd,\Pid^{\S})}}{\pa(x,y,z)}\,.
\eeq

\begin{rem}[The DSI in terms of the potential temperature] As explained in the introduction, the $\DSI$ has been defined by \citet{Nevir2004} in terms of the potential temperature 
\beq
\psi= \theta = \theta_0 \exp\left(\frac{\Sd}{c_{pd}}\right)\,,
\eeq 
\changed{where $\theta_0$ is the potential temperature at reference conditions, for which according to \eqref{theta} the identity $\theta_0=T_0$ holds.} 
This quantity is also conserved during adiabatic frictionless motion because $\theta_0$ and $c_{pd}$ are constants, and so is the related Ertel's potential vorticity
\beq
\Pid^\theta= \frac{\bfxi_a\cdot \nabla\theta}{\rho}\,.
\eeq
With these definitions, and according to \eqref{eq.DSI}, the $\DSI$ as introduced by \citet{Nevir2004} reads 
\beq \label{eq:DSI_d_theta}
\DSId^\theta=\frac{1}{\rho}\nabla \Pid^\theta \cdot(\nabla \theta \times \nabla \Bd)
= \frac{\theta^2}{c_{pd}^2} \DSId^{\S}\,,
\eeq
and it is proportional to the entropy-based version of the $\DSI$ discussed above.
\changed{This in particular means that from the perspective of applications,
where the DSI is used to signal deviations from a balanced state,
these two versions of the DSI are equally good.}
\end{rem}


\section{DSI-Variants with moisture and phase changes }
\label{sec:MoistDSIVariants}


\subsection{The DSI for moist air }
\label{subsec:DSImoist}

Here we derive the Dynamic State Index based on the equations of motion for moist air, but without phase changes, which will be incorporated in section \ref{subsec:DSIc} below. Water components are introduced via the mixing ratios. For water vapor the latter is defined, e.g., as the ratio of the density of water vapor $\r_v$ over the density of dry air $\r_d$,
\beq \label{qv}
q_v=\frac{\r_v}{\r_d} = E \frac{e}{p_d}\,,\qquad \textnormal{where} \quad E= \frac{R_d}{R_v}\,,
\eeq
and where $e$ is the vapor pressure and $R_v$ denotes the ideal gas constant for water vapor, see also table\,\ref{table}. Here we have used that according to the ideal gas law for dry air and water vapor
\beq
p_d=\r_d R_d T\qquad\textnormal{and}\qquad 
e=p_v=\r_v R_v T.
\eeq 
For the total pressure $p$ of moist air (without liquid water) we thus have \changed{according to Dalton's law}
\beq\label{ideal.v1}
p = p_d+e = \r_d (R_d +q_v R_v)T =:\r_d R' T\,,
\eeq
or, equivalently,
\beq\label{ideal.v2}
p = R_d \r T \frac{1+\frac{q_v}{E}}{1+q_v}=: R_d \r T_v\,,
\eeq
where $\r$ is the total density $\rho= \r_d+\r_v$ and $T_v$ is referred to as the virtual temperature, see also \cite{Emanuel1994,Cotton2010}. In a first step, we assume here that $q_v$ is conserved, i.e., we do not allow for phase changes, and that there are no liquid water constituents. This amounts to
\beq
\frac{dq_v}{dt}= \pa_t q_v + \mathbf{v}\cdot \nabla q_v=0\,.
\eeq
Following \citet{Emanuel1994} the total entropy $\Sv$ for the moist air with water vapor only is then given by 
\beq\label{s.v}
\Sv= (c_{pd} +q_v c_{pv}) \ln \frac{T}{T_0} - \Big(R_d\ln \frac{p_d}{p_0} +q_v R_v \ln \frac{e}{e_0}\Big)\,,
\eeq
where $c_{pv}$ denotes the specific heat capacity at constant pressure for water vapor, see also table\,\ref{table}, and 0-indices denote reference values. Making use of \eqref{qv}, the entropy can alternatively be rewritten as 
\beq
\Sv= c_p' \ln \frac{T}{T_0} - R' \ln \frac{p}{p_0} + R'\ln \big(1+\frac{q_v}{E}\big) -q_v R_v \ln\frac{q_v}{q_{v0}}\,, \ \ \ \ \ 
\eeq
where here and below we let 
\beq
c_p'=c_{pd} + q_v c_{pv} \,,\qquad R'= R_d +q_v R_v\,.
\eeq
The differential reads
\beq\label{dsv}
d\Sv=c_p'd\ln T  - R' d  \ln p + \big(c_{pv}\ln \frac{T}{T_0} - R_v \ln \frac{e}{e_0} \big) d q_v\,, \ \ 
\eeq
where we have used $\ln e/e_0=\ln p/p_0 +\ln q_v/q_{v0} - \ln (1+q_v/E)$.
Notice that due to the conservation of $q_v$ we find 
\beq \label{eq:dsdt_ma}
\frac{d \Sv}{dt}= c_p'  \frac{d\ln T}{dt} -  R' \frac{d\ln p}{dt} \,.
\eeq 
During isentropic motion, $\Sv$ is conserved and we choose
\beq
\psi=\Sv\, \qquad \textnormal{and} \qquad \Pi^{s}_{(v)}=\frac{\bfxi_a \cdot\nabla \Sv}{\rho} \,.
\eeq
To obtain the appropriate definition for the steady wind and the DSI for moist air with water vapor it now remains to determine the Bernoulli function $\Bv$ for moist air and accordingly the remainder function $\Gv$. The total moist enthalpy is (up to constants) given by
\beq
\hv=c_p' T\,,
\eeq
satisfying
\beq
d \hv= c_p' dT +c_{pv} T dq_v\,.
\eeq
We set
\beq \label{Hv}
\Hv=  \frac{\hv}{1+q_v} \,,
\eeq 
where we note that the normalization of $\hv$ by $(1+q_v)=\frac{\r}{\r_d}$ accounts for the fact that $\frac{1}{\r_d}\nabla p$ arises in the gradient of $\Hv$, or $T\nabla \Sv$ respectively, instead of $\frac{1}{\r} \nabla p$ appearing in the definition of $\mathbf{G}$, see also \eqref{Gv} below. 

Using \eqref{dsv} and \eqref{ideal.v1} one can compute 
\beq
\frac{1}{\r_d}\nabla p\, &=& \,TR' \nabla \ln p \\ 
&=&c_p'\nabla T -T \nabla \Sv + T\Big(c_{pv} \ln \frac{T}{T_0} - R_v \ln \frac{e}{e_0}\Big)\nabla q_v\nonumber\\
&=&(1+q_v) \nabla \Hv- T \nabla \Sv + (1+q_v)\Lambda_{(v)} \nabla q_v\,,
\eeq
where we denote 
\beq\label{A.v.s}
\Lambda_{(v)}=\frac{1}{1+q_v} \Big(\Hv + c_{pv} T \big(\ln \frac{T}{T_0}-1\big) - R_v T \ln \frac{e}{e_0}\Big)\,,
\eeq
such that by the definition of $\mathbf{G}$ in \changed{\eqref{eq.G}} we have
\beq\label{Gv}
\changed{\Gv=\nabla \Hv - \frac{\nabla p}{\rho} = \frac{1}{1+q_v} T\nabla \Sv - \Lambda_{(v)}\nabla q_v}
\eeq
and for the cross product
\beq
\Gv \times \nabla \Sv = \changed{-\Lambda_{(v)} \nabla q_v\times \nabla \Sv}\,.
\eeq
The Bernoulli function for moist air without phase changes reads 
\beq
\Bv = \frac{1}{2} \mathbf{v}^2 + \Hv+ \phi \,.
\eeq	
Therefore the steady wind results in
\beq
\mathbf{v}_{st,(v)}=- \frac{1}{\rho \Piv }\Big( (\nabla \Bv + \Lambda_{(v)}  \nabla q_v)\times \nabla \Sv\Big)\,.
\eeq
This steady wind describes a basic state that contains moist air without phase changes. Compared to the steady wind for dry air, this basic state $\mathbf{v}_{st,(v)}$ has an additional term that contains the water vapor mixing ratio. Deviations from this basic state are related to the generation of clouds and precipitation. These deviations are captured by the DSI for moist air 
\beq\label{DSI.v}
\DSIv = \frac{(\Piv)^2}{\rho} \nabla \cdot\Big[ \frac{1}{\Piv
} \Big(\nabla \Bv  +\Lambda_{(v)}\nabla q_v\Big) \times \nabla \Sv\Big]\,. \ \ \ \ \ \ \ \ 
\eeq
Comparing the $\DSIv$ to  $\DSId$ for dry air \eqref{eq:DSI_d}, the same additional term as in the steady wind representation is added. 
Noting that one velocity component of this steady wind is directed along the isosurfaces of the mixing ratio of water vapor, the DSI captures deviations from this alignment. The $\DSIv$-signals are similar to the signals of the $\DSId$ for dry air, but indicating in more detail the process of moist air transport: The basic state is characterized by vanishing advection tendencies. The DSI captures deviations from the basic state and thus diagnoses the advection of moisture. Therefore, the  $\DSIv$ for moist air without phase changes captures the formation and dissolving of clouds. While the $\DSId$ for dry air signaled deviations from the adiabatic, inviscid and steady basic state, the difference $\DSId-\DSIv$ can be used to locate local deviations from pure transport of moisture. 

\begin{rem}[The DSI based upon a modified potential temperature]
In the moist air case the total derivative of entropy \changed{is} expressed by that of a modified potential temperature $\theta'$:
\beq \label{eq:dsdt}
\frac{ds_{(v)}}{dt}= c_p'  \frac{d\ln T}{dt} -  R' \frac{d\ln p}{dt} = c_p' \frac{d \ln \theta'}{dt}\,,
\eeq
where
\beq 
\theta'=T \left(\frac{p_0}{p}\right)^{\frac{R'}{c_p'}}\,,
\eeq
see also \citet{Emanuel1994}. Thus, during isentropic motion, $\theta'$ is conserved and we could therefore also choose
\beq
\psi=\theta'\, \qquad \textnormal{and} \qquad \Pi^{\theta'}=\frac{\bfxi_a \cdot\nabla\theta'}{\rho} \,. 
\eeq
We note however that the potential vorticity $\Pi^{\theta'}$  is not a conserved quantity anymore, since $\Sv = F(\theta',q_v)$. The simple structure in \eqref{eq:dsdt} is only obtained for the total derivative due to the conservation of $q_v$, but does not hold for the spatial gradient, which involves additional terms proportional to $\nabla q_v$, see also \eqref{dsv}. Thus the solenoidal term in \eqref{cons.Pi} does not vanish anymore. 
Nevertheless following the steps from above the $\DSI$ based upon $\theta'$ reads
\beq\label{DSI.v.theta}
\DSIv^{\theta'} = \frac{(\Piv^{\theta'})^2}{\rho} \nabla \cdot\Big[ \frac{1}{\Piv^{\theta'}
} \Big( (\nabla \Bv  +\Lambda_{(v)}^{\theta'}\nabla q_v)\times \nabla \theta'\Big)\Big]\, \ \ \ \ 
\eeq
with 
\beq
\Lambda_{(v)}^{\theta'}=   \frac{1}{1+q_v} \Big((c_{pv}-c_{pd})T +  \frac{R_v c_{pd}- R_dc_{pv}}{c_p'}T\ln\frac{p_0}{p}\Big)\,. \ \ \ \ 
\eeq
\end{rem}


\subsection{The DSI for cloudy air}
\label{subsec:DSIc}

Here we account for water vapor and cloud liquid water with mixing ratios
\beq
q_v= \frac{\r_v}{\r_d}\,,\qquad q_c=\frac{\r_c}{\r_d}\,,
\eeq
respectively, and for their phase changes in the equations of motion. This leads us to expressions for the steady wind and DSI for cloudy air. Moreover, cloud liquid water, with density $\r_c$, is assumed to be advected by the mean wind, \ie, not to precipitate. Then the total water amount corresponds to
\beq
q_T=q_v+q_c\,,
\eeq
and the total liquid water amount is  
\beq
q_l=q_c\,.
\eeq
For the moisture components we have the balance laws
\beq
\frac{dq_v}{dt} = -\Src_{cd}\,,\qquad \frac{dq_c }{dt}=  \Src_{cd}\,,
\eeq
where $\Src_{cd}$ denotes the \changed{condensation and evaporation} rate. Obviously, the total amount of moisture is conserved, i.e., 
\beq
\frac{dq_T}{dt}=0\,.
\eeq 
The total density $\r$ is given by $\r= \r_d + \r_v+\r_c$ and, following common approximations, the liquid water content is assumed to not exert any pressure on the air parcels, so that
\beq
p=p_d+p_v=p_d+e=\r_d R' T = \r R_d T \frac{1+\frac{q_v}{E}}{1+q_T},
\eeq
where $R'=R_d+q_v R_v$ as above, \cite{Emanuel1994,Cotton2010}.
The total enthalpy in the presence of liquid water reads (again up to constants)
\beq\label{hc}
\hc=c_p' T\,,
\eeq
see, e.g., \cite{Emanuel1994}, where 
\beq
c_p'=c_{pd} + q_v c_{pv} + q_l c_l\,
\eeq
with $c_l$ the specific heat capacity of liquid water, see table\,\ref{table}. Neglecting the temperature dependence of the specific heat capacities provides a good approximation above the melting point, so that the latent heat of vaporization, $L$, which satisfies 
\beq\label{dL}
dL=(c_{pv}-c_l) dT\,, 
\eeq
becomes linear in the temperature, 
\beq\label{L} 
L=L_0 + (c_{pv}-c_l) T \quad \textnormal{with} \ \ L_0=L(T_0)-(c_{pv}-c_l) T_0\,. \ \ \ \
\eeq
Further, the entropy in the case of cloudy air  is given by
\beq\label{s.c}
\Sc= c_p'\ln \frac{T}{T_0} -R_d \ln \frac{p_d}{p_0} -q_v R_v \ln \frac{e}{e_0}. 
\eeq
As in previous steps, we additionally introduce
\beqs
\Hc=\frac{\hc}{1+q_T}\,.
\eeqs
For the gradients we then obtain
\beq
\nabla \hc=c_p' \nabla T +T(c_{pv}\nabla q_v+c_l\nabla q_l)
\eeq
and 
\beq
\nabla \Hc=\frac{\nabla \hc}{1+q_T}-\frac{\Hc}{1+q_T}\nabla q_T\,.
\eeq
Making use of the ideal gas laws for dry air and water vapor, for the entropy accordingly, we obtain
\begin{equation}
\begin{split}
\nabla \Sc = 
  & c_p'\nabla \ln T - \frac{1}{T}\frac{\nabla p}{\r_d} 
    \\
  & + \ln \frac{T}{T_0}(c_{pv} \nabla q_v  + c_l \nabla q_l )  
    - R_v \ln \frac{e}{e_0}\nabla q_v\,,
\end{split}
\end{equation}
and multiplication by $T \r_d/\r=T/(1+q_T)$ yields
\begin{equation}
\begin{split}
\frac{\nabla p}{\r} = 
  & -\frac{T\nabla\Sc}{1+q_T} + \frac{c_p'\nabla T}{1+q_T} 
    \\ 
  & + \big(c_{pv}T \ln\frac{T}{T_0} -R_v T \ln \frac{e}{e_0} \big)\nabla q_v  
    + c_lT \ln \frac{T}{T_0}\nabla q_c\,,
\end{split}
\end{equation}
such that for $\Gc$ we obtain 
\beq
\Gc&=&\changed{\nabla \Hc-\frac{\nabla p}{\r} }
\nonumber\\
&=& 
 \changed{\frac{T\nabla\Sc}{1+q_T} - \Lambda_{(c),1}\nabla q_v  - 
 \Lambda_{(c),2}\nabla q_l\,,}
\eeq
where
\beq\label{Lambdac}
\Lambda_{(c),1}&=& \frac{1}{1+q_T}\Big(\Hc + c_{pv}T \big( \ln \frac{T}{T_0}-1\big) - R_v T \ln \frac{e}{e_0}  \Big)\,,\ \  \\
\Lambda_{(c),2}&=& \frac{1}{1+q_T}\Big(\Hc + c_{l}T \big( \ln \frac{T}{T_0}-1\big)\Big)\,,
\eeq
resemble \eqref{A.v.s}. Again we choose 
\beq
\psi=\Sc\qquad \textnormal{and}\qquad \Pi_{(c)}=\frac{\bfxi_a\cdot \nabla \Sc}{\rho}\,.
\eeq
\changed{Furthermore,}
\beq
\Gc \times \nabla \Sc =\changed{-( \Lambda_{(c),1}\nabla q_v +\Lambda_{(c),2}\nabla q_l)   \times \nabla \Sc}\,,
\eeq
while the Bernoulli function for cloudy air reads
\beq
\Bc= \frac{1}{2}\mathbf{v}^2+ \Hc + \phi\,.
\eeq
Finally, the steady wind results in
\beq
\mathbf{v}_{st,(c)}= \frac{1}{\rho \Pic }\Big[  \nabla \Sc  \times (\nabla \Bc  + \Lambda_{(c),1}\nabla q_v +  \Lambda_{(c),2} \nabla q_l) 
\Big]\,. \ \ \ \ \ \ \ \ 
\eeq
The steady wind for cloudy air describes an atmospheric basic state that includes water vapor, liquid water and phase changes, but no precipitation, and implies the 
\begin{equation}\label{DSI.c.s}
\begin{split}
\DSIc = \frac{(\Pic)^2}{\rho} \nabla \cdot\Big[ \frac{1}{\Pic} \Big( &  (\nabla \Bc + \Lambda_{(c),1}\nabla q_v \\ & +  \Lambda_{(c),2}\nabla q_l)\times \nabla \Sc\Big)\Big]\,.
\end{split}
\end{equation}
Thus, compared to the previously derived $\mathbf{v}_{st,(v)}$ and $\DSIv$ for moist 
air without phase changes the $\DSIc$ for cloudy air is extended by a term proportional 
to the gradient of the liquid water content, and latent heating appears explicitly 
through the definition of $\Sc, \Bc$ and $\Lambda_{(c),i}$, respectively. The utility
of this new index is that the difference $\DSIc - \DSIv$ indicates processes associated
with cloud generation or dissolution.

In the next section, we will further include the (vertical) transport of precipitation 
to derive a DSI variant that signals, e.g., extreme or less intense precipitation. 

\begin{rem}[Alternative expressions for the thermodynamic quantities]
Often the enthalpy in the case of liquid water being present is stated as (again up 
to constants)
\beqs
\hc=\changed{(c_{pd} + q_T c_l)} T+ Lq_v
\eeqs
see, e.g., \cite{Emanuel1994}. Using \eqref{L}, this coincides with \eqref{hc} up to 
the constant value $L_0$. Accordingly, a common formulation of the entropy for cloudy 
air is
\beq\label{sc1}
\Sc= (c_{pd}+q_T c_l)\ln \frac{T}{T_0} -R_d \ln \frac{p_d}{p_0}+ \frac{L q_v}{T} -q_v R_v \ln \frac{e}{e^*}\,.\ \ \
\eeq 
According to \cite{Emanuel1994}, the latent \changed{heat} $L$ satisfies  
\beq\label{LT}
\frac{L}{T}=(c_{pv}-c_l)\ln \frac{T}{T_0} - R_v \ln \frac{e^*}{e_0}\,,
\eeq
which is consistent with \eqref{dL} owing to the Clausius-Clapeyron relation for 
the saturation vapor pressure $e^*$, 
\beq\label{CC}
d\ln e^*= \frac{LdT}{R_vT^2}\,.
\eeq 
This verifies, in particular, the equivalence of \eqref{sc1} and \eqref{s.c}.
\end{rem}


\subsection{The DSI for precipitating air}
\label{subsec:DSIr}

To cover the precipitation of rain, the ``rain amount'' 
\beq
q_r=\frac{\r_r}{\r_d}\,,
\eeq
is introduced and its evolution includes vertical sedimentation with the terminal
fall velocity, $V_r$. Then the total water amount and total density correspond to
\beq\label{qT.r}
q_T=q_v+q_c+q_r\,, \qquad \r= \r_d+\r_v+\r_c+\r_r\,,
\eeq
and the liquid water content is determined by 
\beq\label{ql.r}
q_l= q_c+q_r\,.
\eeq  
The moisture quantities satisfy the balance laws
\beq
\frac{dq_v}{dt} &=& -\Src_{cd} + \Src_{ev}\,,\\
 \frac{dq_c}{dt}&=&  \Src_{cd}-\Src_{ac}-\Src_{cr}\,, \\
 \frac{dq_r}{dt}- \frac{1}{\rho_d}\pa_z (\r_d V_r q_r)&=&  -\Src_{ev}\changed{+}\Src_{ac}\changed{+}\Src_{cr}\,,
\eeq
where $\Src_{cd}$ again denotes \changed{condensation and evaporation} rates, $\Src_{ev}$ the 
evaporation rate of rain, $\Src_{ac}$ the autoconversion rate of cloud water into rain 
once droplets grow big enough, and $\Src_{cr}$ is the collection rate of cloud water by 
the falling rain. The total amount of moisture is conserved up to the relative vertical 
transport of precipitation, so that
\beq
\frac{dq_T}{dt}=\frac{1}{\rho_d}\pa_z (\r_d V_r q_r)\,.
\eeq 
The terminal fall velocity of precipitation affects also the momentum balance \eqref{eq.v}, 
which is extended by an additional term  
\beq
\mathbf{W}_r=\frac{1}{1+q_T}\pa_z (q_r V_r \mathbf{v})\,
\eeq
on the right hand side, see \cite{Cotton2010}, \cite{HittmeirKlein2018}, which is typically 
neglected in the literature. In an asymptotic analysis for deep convective clouds it 
was found not to play a role in the leading order dynamics on shorter time scales, but 
the term could get to be relevant for longer time scales or over large areas of 
precipitation, \cite{HittmeirKlein2018}. This is also in agreement with \cite{Cotton2010} and 
references therein. As we are interested here in analysing \changed{warm} convective events and local 
processes, it is acceptable to assume 
\beq
\mathbf{W}_r=0\,,
\eeq
below. Then the steady wind can be derived in a similar fashion as before and, 
following the earlier derivations for cloudy air by replacing $q_T$ with \eqref{qT.r} 
and $q_l$ with \eqref{ql.r}, we obtain the same expressions for 
$\Br, \Gr, \Hr,\Pir,\Lambda_{(r),i}$ as for $\Bc,\Gc,\Hc,\Pic, \Lambda_{(c),i}$. 
The precipitation terms, however, affect the entropy balance by constituting a source 
term 
\beq
\frac{d\Sr}{dt}=c_l \ln \frac{T}{T_0}\frac{dq_T}{dt}=c_l \ln \frac{T}{T_0}\frac{1}{\rho_d}\pa_z (\r_d V_r q_r) =: \Src_{\Sr}\,, \ \ \ \ \ \ 
\eeq
and the generalised construction of the steady wind yields
\beq
\mathbf{v}_{st,(r)}&=&
 \frac{1}{\rho \Pir }\Big[  \nabla \Sr  \times (\nabla \Br  + \Lambda_{(r),1} \nabla q_v+  \Lambda_{(r),2} \nabla q_l) + \nonumber\\
&&\qquad \qquad  + \bfxi_a\,  \Src_{\Sr}
\Big] \,.
\eeq
Thus, for the $\DSIr$ we have
\begin{equation}\label{DSI.c.s}
\begin{split}
\DSIr  =&  \frac{(\Pir)^2}{\rho} \nabla \cdot\Big[ \frac{1}{\Pir
} \Big( (\nabla \Br + \Lambda_{(r),1}\nabla q_T + \\ 
 & \qquad \qquad +  \Lambda_{(r),2}\nabla q_v)\times \nabla \Sr - \bfxi_a\,  \Src_{\Sr} \Big)\Big]\,.
\end{split}
\end{equation}
\changed{Relative to the DSI for dry air in \eqref{eq:DSI_d}, the $\DSIr$ indicates deviations 
from a basic state to the balance of which both phase changes of water and the 
(vertical) transport of precipitation contribute substantially. In turn, comparisons of 
the $\DSIr$ with the moist and cloudy air variants, $\DSIv$ and $\DSIc$, respectively, 
allows the user to distinguish regions in which different combinations of these processes 
balance each other.}


\section{Conclusion}
\label{sec:Concl}

For the non-hydrostatic compressible governing equations without moist processes, non-zero
values of the scalar dynamic state index, DSI, indicate non-stationary, diabatic and 
dissipative atmospheric processes. This work generalized this concept to moist atmospheric
flows, ultimately including phase changes and precipitation. The point of departure for
the present developments is the observation that the original dry air \DSI\ has a 
representation in terms of the mass flow divergence of Schär's \cite{Schaer1992} 
``steady wind'', $\mathbf{v}_{st}$,
\beq\label{eq:DSIconcl}
\DSI \equiv \frac{1}{\rho} \frac{\partial(\theta,B,\Pi^\theta)}{\partial(x,y,z)}
     = -\frac{\Pi^2}{\rho} \nabla \cdot (\rho \mathbf{v}_{st}^\theta)\,.
\eeq
It is difficult to see how the determinant of gradients of the constitutive Lagrangian 
conserved quantities $(\theta, B, \Pi^\theta)$, i.e., of potential temperature, 
Bernoulli function, and potential vorticity, respectively, can be generalized to 
thermodynamically more complex situations. In contrast, generalization of the concept
of the steady wind has turned out to be accessible, and has allowed us to achieve the
stated goal.

Thus, in a hierarchical fashion we have introduced three generalizations to include 
moist processes in the DSI-concept. First, we included water vapor neglecting phase 
changes. Second, we considered water vapor together with phase changes to account for 
cloud formation, and, third, we have included the fall out of precipitation. For all 
models we first derived the associated steady wind representing the basic state, noticing 
that only the basic state for moist air without phase changes still characterizes 
adiabatic conditions. The second generalization of the basic state incorporates diabatic 
but also reversible processes. With these preliminaries, the DSI is given by expressions
analogous to the last term in \eqref{eq:DSIconcl}, which can be transferred to all models 
once appropriate analoga to the \changed{potential vorticity $\Pi$} and the steady wind 
$\mathbf{v}_{st}$ are found. Stepwise extensions of the underlying flow models by different 
moist process descriptions leads to a hierarchy of steady wind representations, for 

\noindent dry air 
\begin{equation} 
 \mathbf{v}_{st,(d)} = \frac{1}{\rho \changed{\Pid} } 
\Big[  \nabla \Sd  \times \nabla \Bd   \Big]\,, 
\end{equation}
moist air
\begin{equation} 
\mathbf{v}_{st,(v)} = \frac{1}{\rho \Piv }
\Big[ \nabla\Sv\times (\nabla \Bv + \Lambda_{(v)}  \nabla q_v)\Big] \,,
\end{equation}
cloudy air
\begin{equation}
 \begin{split}
 \mathbf{v}_{st,(c)} & = \frac{1}{\rho \Pic } 
\Big[  \nabla \Sc  \times (\nabla \Bc  + \Lambda_{(c),1} \nabla q_v +\\ 
& \hspace{3.5cm} +  \Lambda_{(c),2}\nabla q_l)  \Big] \,, 
\end{split}
\end{equation}
fall out of precipitation
\begin{equation}
 \begin{split}
\mathbf{v}_{st,(r)} & = \frac{1}{\rho \Pir }
\Big[  \nabla \Sr  \times (\nabla \Br  + \Lambda_{(r),1} \nabla q_v +\\
& \hspace{2.5cm}+  \Lambda_{(r),2} \nabla q_l)  + \bfxi_a\,  \Src_{\Sr} \Big]\,,
\end{split}
\end{equation}
where the density $\rho$ and the pressure $p$ denote the total density and total pressure 
for each of the different aerodynamic models. Comparing the dry air case, where the steady 
wind blows parallel to level sets of the Bernoulli function on isentropic surfaces, with 
the moist air case, an additional velocity component appears that is directed along the 
isolines of the mixing ratio of water vapor. If phase changes take place, liquid water 
needs to be accounted for, too, which generates an additional contribution to the 
steady wind. The latent heat then arises in the definition of the entropy and potential 
vorticity.


\begin{table*}
	\caption{\label{tab:DSIsummary}Physical processes and characterizations, the advections of the entropy and the PV with respect to the steady wind vanish, if they are conserved.}
		\begin{tabular}{llll}
			\hline
			 & entropy & PV   & Specific DSI signals \\
			\hline\hline
			  \textbf{Dry air} & $s_{(d)}$ &  $\Pi _{(d)}  $ &  $\DSId  \neq 0$    \\
			   & conserved & conserved & all diabatic (frictional, non-steady)    \\
			   & no advection & no advection & processes    \\
			   \hline					
 \textbf{Moist air} & $s_{(v)}$ &  $\Pi _{(v)}  $ &  $\DSIv - \DSId  \neq 0$     \\
			   & conserved & conserved &  transport of moist air \\ 
			  & no advection & no advection &  variations of the humidity 			  \\
\hline
 \textbf{Cloudy air} & $s_{(c)}$ &  $\Pi _{(c)} $ &  $\DSIc - \DSIv   \neq 0$     \\
			   & conserved & conserved & the generation and dissolving of clouds  \\
			     & no advection & no advection &    			     \\
			   \hline
 		 \textbf{Fallout of rain} & $s_{(r)}$ &  $\Pi _{(r)} $ & $ \DSIr -\DSIc \neq 0$     \\
			    & not conserved & not conserved & variations of precipitation  \\
			   & advection  & advection & (e.g. in form of ice)  \\
			 \hline
 		\end{tabular}
\end{table*}


With these results, the new $\DSI$ for moist aerothermodynamics results from the respective
generalizations of steady wind mass flux divergence term on the right of \eqref{eq:DSIconcl}. 
For conservative systems the $\DSI$ can still equivalently be formulated based upon 
Jacobian-determinants:
\begin{align}
\DSId 
  & = \frac{1}{\r_d}\nabla \Pid\cdot (\nabla \Sd\times \nabla \Bd) 
    = \frac{1}{\rho} \frac{\partial (s,B,\Pi)}{\partial (x,y,z)} 
      \nonumber\\
  & = \frac{\Pid^2}{\rho} \nabla \cdot\Big[ \frac{1}{\Pid} 
      \Big( \nabla \Bd\times \nabla \Sd \Big)\Big] \, ,\medskip
    \\
\DSIv 
  & = \frac{\Piv^2}{\rho} \nabla \cdot\Big[ \frac{1}{\Piv} 
      \Big( (\nabla \Bv+\Lambda_{(v)}\nabla q_v)\times \nabla \Sv\Big)\Big]
    \nonumber\\ 
  & = \frac{1}{\rho} \Bigg( \frac{\partial (\Sv,\Bv,\Piv)}{\partial (x,y,z)} 
    - \Lambda_{(v)} \frac{\partial (\Piv,q_v,\Sv)}{\partial (x,y,z)}  \nonumber 
    \\ 
  & \hspace{1cm} - \Piv \frac{\partial (\Lambda_{(v)},q_v,\Sv)}{\partial(x,y,z)} \Bigg)  \, ,
    \medskip\\
\DSIc
  & = \frac{\Pic^2}{\rho} \nabla \cdot\Big[ \frac{1}{\Pic} \big( \nabla \Bc + \Lambda_{(c),1}\nabla q_v +  
    \nonumber\\
  & \hspace{3cm} + \Lambda_{(c),2}\nabla q_c\big)\times \nabla \Sc \Big] \, ,
    \medskip\\
\DSIr
  & = \frac{\Pir^2}{\rho} \nabla \cdot 
      \Big[ \frac{1}{\Pir} \Big( \big(\nabla \Br + \Lambda_{(r),1}\nabla q_v +
    \nonumber\\
  & \hspace{2cm} +  \Lambda_{(r),2}\nabla (q_c+q_r)\big)\times \nabla \Sr -\bfxi_a\,  \Src_{\Sr}\Big) \Big] \, .
\end{align}
Reformulating accordingly $\DSIc$ as a sum of five Jacobian-determinants is straightforward. 
Only in the presence of precipitation, with the Lagrangian conservation of the relevant 
constituents of the $\DSI$ no longer guaranteed, can the $\DSI$ no longer be written solely 
in terms of Jacobi-determinants. \changed{For convenience we have summarized all formulae needed to
directly evaluate the various DSI-variants in Appendix~A.}

For the derivations in this paper we have preferred using the entropy as the relevant 
advected scalar and in formulating a potential vorticity variable over any one of various
possible potential temperatures. Our motivation is that the entropy has a unique physical 
meaning across all cases, which is generally not true for potential temperatures in complex 
multicomponent flows. This is not to say, however, that a particular choice of
a potential temperature variable could not streamline some of the derivations or have 
advantages in terms of physical interpretability. For example, by adopting 
the purely pressure and density dependent potential temperature $\theta_\rho$ in 
\cite{Schubert2004} for the formulation, one can enforce the potential vorticity to remain
a Lagrangian conserved quantity even in the presence of precipitation, and this may help
interpretations or further in-depth analyses. The recent study by \citet{Baumgartner2020}, 
who investigates the potential temperature in terms of temperature dependent specific heat 
capacities $c_p(T)$, might also be of interest in this context, and it could be incorporated 
in the present framework as well.

The $\DSId$ for dry air, $\DSIv$ for moist air, the $\DSIc$ for cloudy air and the $\DSIr$ 
for precipitating air indicate deviations of local flow conditions from inviscid and steady 
state motions. Differences of $\DSI$ variants that encode different steady balances can be
utilized to identify and locate particular diabatic processes. While the classical $\DSId$ 
for dry air reflects general diabatic processes, the basic state of $\DSIv$ additionally 
contains water vapor. Thus the difference $\DSId-\DSIv$ indicates the transport of moist air. 
Considering cloudy air, \changed{\ie} adding the effects of liquid water and its phase changes to the 
equations of motion, and thus to the basic state, the $\DSIc$ for cloudy air indicates 
general precipitation processes and other diabatic processes, such as radiation. The according 
difference $\DSIv-\DSIc$ therefore signals the generation and dissolution of clouds. Accounting,
in addition, for the (vertical) transport of precipitation, the difference $\DSIr -\DSIc$ 
acts as an indicator for the occurrence and intensity of rain. The different $\DSI$ variants 
and their proper physical interpretations are summarized in table~\ref{tab:DSIsummary}. 

An interesting further challenge will be the incorporation of the ice phase and its different
conformations, such as snow, graupel, or hail.


\section*{Acknowledgments}
SH thanks the Austrian Science Fund (FWF) for the support via
the Hertha-Firnberg project FWF:~T-764 \changed{and via the SFB project 'Taming Complexity in Partial Differential Equations' FWF: F65}. 
This research has been partially funded by Deutsche Forschungsgemeinschaft (DFG) through grant CRC 1114 'Scaling Cascades in Complex Systems, Project Number 235221301, Projects A01 'Coupling a multiscale stochastic precipitation model to large scale atmospheric flow dynamics' and C06 'Multiscale structure of atmospheric vortices'.

Data sharing is not applicable to this article as no new data were created or analyzed in this study.


\appendix


\section*{Summary of the DSI variants}

In the following we summarise all DSI variants and highlight the additional terms entering when moving up the hierarchy of complexity involved in the according derivations. \changed{We shall emphasize that $\Delta$ in the following does not denote the Laplacian, but is used as a symbol for the deviation terms.} 


\section{Dry air ($\DSId$)}

For the derivations based upon the dry air setting we have  the classical definition of the DSI
\beqs
\DSId&=&\frac{1}{\r_d}\nabla \Pid\cdot (\nabla \Sd\times \nabla \Bd)\\
&=&\frac{\Pid^2}{\r_d}\nabla \cdot \left[\frac{1}{\Pid}\big(\nabla \Bd\times \nabla \Sd\big)\right]\,,
\eeqs
where 
\beqs
\Sd&=&c_{pd} \ln \frac{T}{T_0} - R_d \ln \frac{p}{p_0}\\
\Hd&=& \ c_{pd} T\\
\Bd&=&\frac12 \mathbf{v}^2 + \Hd +\phi\\
\Pid&=&\frac{\boldsymbol{\xi_a}\cdot \nabla \Sd}{\r_d} \, .
\eeqs


\section{Moist air ($\DSIv$)}
As a next step water vapor is included into the derivations, leading to:
\begin{align*}
\DSIv&=\frac{\Piv^2}{\r}\nabla \cdot \left[\frac{1}{\Piv}(\nabla \Bv + \Lambda_{(v)} \nabla q_v)\times \nabla \Sv \right]\\
&=\frac{(\Pid + \Delta \Piv)^2}{\r_d(1+ q_v)}\nabla \cdot \Bigg[\frac{1}{\Pid +\Delta\Piv}\big(\nabla (\Bd + \Delta \Bv ) \\
& \hspace{3cm} + \Lambda_{(v)} \nabla q_v\big)\times \nabla (\Sd + \Delta \Sv)\Bigg] \, , \nonumber
\end{align*}
where 
\begin{align*}
\Sv&=\Sd + \Delta \Sv\,,\\  
\Delta \Sv &= c_{pv} q_v \ln \frac{T}{T_0} - R_v q_v \ln \frac{e}{e_0} \, , \\
\Hv&=\Hd + \Delta \Hv\,, \\
\Delta \Hv &=-\frac{q_v\Hd}{1+q_v}+\frac{c_{pv} q_v T}{1+q_v} \, ,\\
\Bv&=\Bd+ \Delta \Bv\,, \\
\Delta \Bv &= \Delta \Hv \, ,\\
\Piv&=\Pid + \Delta \Piv\,, \\
\Delta \Piv &=-\frac{q_v\Pid}{1+q_v} + \frac{\boldsymbol{\xi_a}\cdot \nabla \Delta \Sv}{\r_d(1+q_v)}\,,\\
\Lambda_{(v)}&=\frac{1}{1+q_v} \Big(\Hv + c_{pv} T \big(\ln \frac{T}{T_0}-1\big) - R_v T \ln \frac{e}{e_0}\Big)\ \, .
\end{align*}


\section{Cloudy air ($\DSIc$)}

The next extension amounts to the inclusion of phase changes and liquid water in the form of cloud water:
\begin{equation*}
\begin{split}
\DSIc&= \frac{\Pic^2}{\rho} \nabla \cdot\Big[ \frac{1}{\Pic} \big( \nabla \Bc + \Lambda_{(c),1}\nabla q_v +  \Lambda_{(c),2}\nabla q_c\big)\times \nabla \Sc \Big]\\
&=\frac{(\Piv + \Delta \Pic)^2}{\r_d(1+ q_v+ q_c)}\nabla \cdot \Big[\frac{1}{\Piv +\Delta\Pic}\Big(\big(\nabla (\Bv + \Delta \Bc ) +\\
& + (\Lambda_{(v)} + \Delta\Lambda_{(c),1}) \nabla q_v + \Lambda_{(c),2}\nabla q_c \big)\times \nabla (\Sv +\Delta \Sc)\Big)\Big]\,, \nonumber
\end{split}
\end{equation*}
where 
\begin{equation*}
\begin{split}
\Sc&=\Sv + \Delta \Sc\,, \\
\Delta \Sc &= c_lq_c\ln \frac{T}{T_0}\,, \\
\Hc&=\Hv + \Delta \Hc\,,  \\
\Delta \Hc &=-\frac{q_c \Hv}{1+q_v+q_c}+\frac{c_lq_cT+L_0}{1+q_v+q_c}\,, \\
\Bc&=\Bv+ \Delta \Bc \,, \\
\Delta \Bc &= \Delta \Hc \,,\\
\Pic&=\Piv + \Delta \Pic\,, \\
\Delta \Pic &=-\frac{q_c \Piv}{1+q_v+q_c} + \frac{\boldsymbol{\xi_a}\cdot \nabla \Delta \Sc}{\r_d(1+q_v+q_c)}\,,\\
\Lambda_{(c),1}& =\Lambda_{(v)} + \Delta \Lambda_{(c),1}\,,   \\
\Delta \Lambda_{(c),1}&= - \frac{q_c \Lambda_{(v)}}{1+q_v+q_c}  + \frac{\Delta \Hc}{1+q_v+q_c}\,,\\
 \Lambda_{(c),2} & =\frac{1}{1+q_v+q_c} \Big(\Hc + c_{l}T \big( \ln \frac{T}{T_0}-1\big) \Big)\,.
\end{split}
\end{equation*}


\section{Fallout of rain ($\DSIr$)}
Finally, also precipitation is included into the derivation of the DSI:
\begin{equation*}
\begin{split}
&\DSIr= \frac{\Pir^2}{\rho} \nabla \cdot\Bigg[ \frac{1}{\Pir} \Big( \big(\nabla \Br + \Lambda_{(r),1}\nabla q_v \\ 
&  \hspace{2cm}+  \Lambda_{(r),2}\nabla (q_c+q_r)\big)\times \nabla \Sr -\bfxi_a\,  \Src_{\Sr}\Big) \Bigg]\\
&=\frac{(\Pic + \Delta \Pir)^2}{\r_d(1+ q_v+ q_c+ q_r)}\nabla 
\cdot \Bigg[\frac{1}{\Pic +\Delta\Pir}\Big(\big(\nabla (\Bc + \Delta \Br ) \\ 
&  \quad + (\Lambda_{(c),1} + \Delta\Lambda_{(r),1}) \nabla q_v + \dots \\ 
& \quad + (\Lambda_{(c),2}+\Delta\Lambda_{(r),2})(\nabla q_c+ \nabla q_r)\big) 
 \times \nabla (\Sc +\Delta \Sr) \\
 & \qquad -\bfxi_a\,  \Src_{\Sr} \Big)\Bigg]\,, \nonumber
\end{split}
\end{equation*}
where 
\begin{align*}
\Sr&=\Sc + \Delta \Sr\,,\\ 
\Delta \Sr &= c_lq_r\ln \frac{T}{T_0}\,, \\
\Hr&=\Hc + \Delta \Hr\,,\\  
\Delta \Hr &=-\frac{q_r\Hc}{1+q_v+q_c+q_r}+\frac{c_lq_rT}{1+q_v+q_c+q_r} \,,\\
\Br&=\Bc+ \Delta \Br \,,\\ 
\Delta \Br &= \Delta \Hr \,,\\
\Pir&=\Pic + \Delta \Pir\,,\\ 
\Delta \Pir&=-\frac{q_r\Pic}{1+q_v+q_c+q_r} + \frac{\boldsymbol{\xi_a}\cdot \nabla \Delta \Sr}{\r_d(1+q_v+q_c+q_r)}\,,\\
\Lambda_{(r),1}&=\Lambda_{(c),1} + \Delta \Lambda_{(r),1}\,, \\
\Delta   \Lambda_{(r),1} &=  - \frac{q_r \Lambda_{(c),1}}{1+q_v+q_c+q_r}  + \frac{\Delta \Hr}{1+q_v+q_c+q_r}\,,\\
\Lambda_{(r),2}&=\Lambda_{(c),2} + \Delta \Lambda_{(r),2}\,, \\
\Delta   \Lambda_{(r),2} &=  - \frac{q_r \Lambda_{(c),2}}{1+q_v+q_c+q_r}  + \frac{\Delta \Hr}{1+q_v+q_c+q_r}\,,\\
\Src_{\Sr}&=c_l \ln \frac{T}{T_0}\frac{1}{\rho_d}\pa_z (\r_d V_r q_r) \,.
\end{align*}

\end{document}